\documentclass[reprint,aps,twocolumn,superscriptaddress,pra]{revtex4-2}

\usepackage{amsmath}
\usepackage{amssymb}
\usepackage{color}
\usepackage{graphicx}
\usepackage{svg}
\usepackage{bm}      
\usepackage{url}
\usepackage[bookmarks=false,colorlinks,citecolor=blue]{hyperref}
\usepackage{comment}
\usepackage{braket}

\begin{document}

\title{Stabilization of bulk quantum orders in finite Rydberg atom arrays}

\author{Yash M. Lokare}
\affiliation{Department of Physics, Brown University, Providence, Rhode Island 02912, USA
}%
\author{Matthew J. Coley-O'Rourke}
\affiliation{Department of Chemistry, Brown University, Providence, Rhode Island 02912, USA
}%

\date{\today}

\begin{abstract}
Arrays of ultracold neutral atoms, also known as Rydberg atom arrays, are rapidly developing into a powerful and versatile platform for quantum simulation. 
However, theoretical predictions about the bulk quantum phases of matter present in these systems have often diverged from experimental realizations on finite-sized arrays due to the strong effects of the boundaries.
Here we propose a general, experimentally straightforward strategy to mitigate the effects of the boundaries and thus enable finite-sized arrays to stabilize bulk-like quantum order. Our scheme makes use of the properties of the ubiquitous disordered phase in Rydberg systems, driving the boundaries into an unbiased set of configurations that depend on the bulk physics. 
We numerically demonstrate the efficacy of this protocol in one- and two-dimensional systems on both ordered and critical phases.
\end{abstract}

\maketitle

Programmable arrays of Rydberg atoms are currently a leading platform for quantum simulation and computation. They combine strong, tunable interactions, native geometrical flexibility, and high-fidelity measurement to enable experimental access to a broad range of quantum many-body phenomena. These include quantum spin models~\cite{Browaeys2016, Scholl2021, Kim2024}, constrained dynamics and quantum scars \cite{Bernien2017, Bluvstein2021, Manovitz2025, Datla2025}, quantum optimization protocols \cite{Ebadi2022}, and complex symmetry-broken and topologically-ordered quantum phases \cite{Ebadi2021,Semeghini2021, Kanungo2022, Evered2025, Bornet2026}.
In addition to these various experimental demonstrations, myriad theoretical proposals further suggest a broad range of exciting physical phenomena that could be realized using this platform \cite{Myerson2022, Giudici2022, Shen2023, Giudici2023, Macedo2024, Vovrosh2024, Euchner2025, Koyluoglu2025}. 

Despite this rapid progress, 
a persistent issue remains throughout the literature: the results from exquisitely controlled experiments consistently deviate both quantitatively and qualitatively from theoretical predictions~\cite{Samajdar2020,Ebadi2021,Verresen2021,Semeghini2021,Rader2019,Zhang2025,Samajdar2024course,Manovitz2025,Hirsbrunner2025}. This discrepancy is often attributed to ``finite size effects,'' a consequence of theoretical predictions made for periodic or infinite systems while the experiments are necessarily finite-sized. Recent results have shown that these effects are strong even on large lattices consisting of $200+$ atoms~\cite{Kalinowski2022, ORourke2023, Hirsbrunner2025}, and that simply enlarging the size of the experiment does not easily mitigate the important effects~\cite{Zhang2025}. 

The sensitivity of Rydberg arrays to boundary conditions is not merely a technical inconvenience but a fundamental aspect of strongly interacting lattice systems. 
The impact of boundary effects is a direct result of the strong interactions present in Rydberg atom systems, which is precisely the same property that also makes them effective platforms for quantum science. This renders typical simulation strategies for reducing finite size effects, such as enlarging the system size or modifying the interactions, experimentally ineffective or undesirable. 

Motivated by the need to clarify and overcome these issues, this work proposes a general, physically inspired, and experimentally straightforward technique to suppress boundary effects in finite Rydberg arrays. It relies only on local control of the atomic Hamiltonian~\cite{Chen2023, Bornet2024, Manovitz2025, Oliveria2025} and makes use of the intrinsic properties of the well-known disordered (or, ``paramagnetic'') phase, which is stabilized in typical Rydberg atom arrays regardless of dimension and geometry~\cite{Low2009, Weimer2010, Browaeys2020}.

The Hamiltonian for an array of interacting neutral atoms individually trapped and coherently driven from their ground state $\ket{g}$ to a Rydberg state $\ket{r}$ is given by,
\begin{equation}
    \label{rydberg hamiltonian}
    H = \displaystyle\sum_{i} \biggl[\frac{\Omega}{2} (\ket{g_i}\bra{r_i} + \ket{r_i}\bra{g_i}) - \delta_i \hat{n}_i\biggr] + \displaystyle\sum_{i < j} \frac{R_b^6}{|(\mathbf{x}_i - \mathbf{x}_j)|^6} \hat{n}_i \hat{n}_j.
\end{equation}
Here $\Omega$ denotes the Rabi frequency, $\delta_i$ is the (possibly site-dependent) detuning from resonance, $\hat{n}_i = \ket{r_i}\bra{r_i}$, and $i$, $j$ label sites at positions $\mathbf{x}_{i}$ of the lattice. The variable interaction strength is characterized by a mutual excitation blockade radius $R_b$. Working in units of $\Omega = 1$ yields two tunable parameters $\delta$ and $R_b$~\cite{Samajdar2020}. We study this Hamiltonian using large-scale density matrix renormalization group (DMRG) simulations \cite{White1992, Schollwock2005}, as implemented in the \texttt{BLOCK2} simulation package~\cite{Zhai2023}, retaining all long-range interaction terms without truncation (see Supplemental Materials (SM) \cite{Supplementary} for details).

\begin{figure}[ht]
    \includegraphics[width=\linewidth]{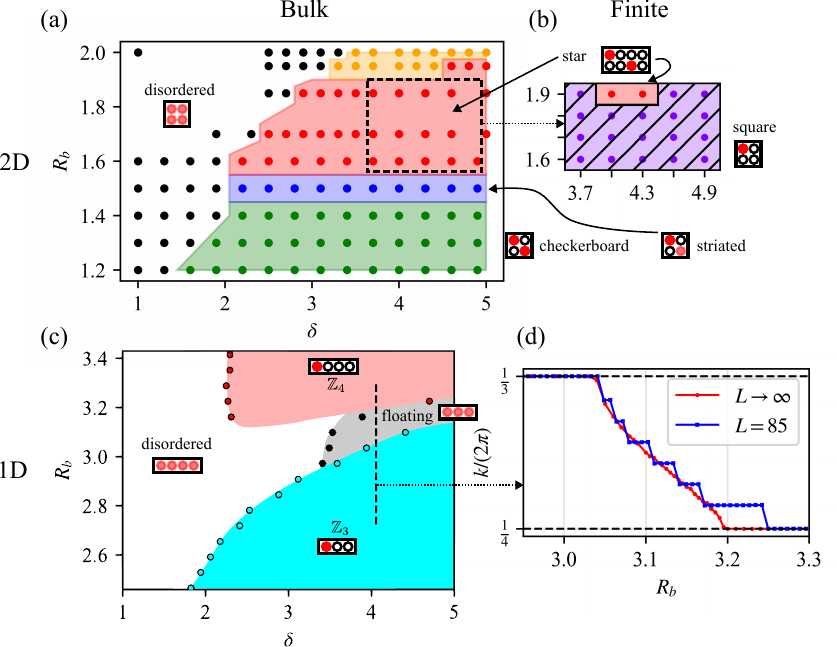}
    \caption{Strong influence of boundary effects. Ground state phase diagrams in the thermodynamic limit (or, ``bulk'') are shown for (a) 2D square lattice and (c) 1D chain geometries. Color coding indicates different phases and insets show the real-space density-wave order (red corresponds to $\ket{r}$, white to $\ket{g}$).
    The ground state behavior on corresponding finite lattices is shown for (b) $13 \times 13$ and (d) $L=85$ systems. (b) In 2D, the square order ground state (hatched purple) is stabilized in a parameter regime where the thermodynamic-limit reference is the star phase. (d) In 1D, a line cut at $\delta = 4.06$ shows the dominant wavevector $k$ of the spatial fluctuations in $\langle n_i \rangle$ for lattices with $L = 85$ and $L = 1009$ sites.
     }
     \label{TL behavior 1D and 2D}
\end{figure}

The competition between coherent driving, detuning, and interaction strength gives rise to a rich set of ground state quantum phases in both one ($1$D) and two ($2$D) dimensions. Recent studies have primarily focused on ordered phases and spatially uniform Hamiltonians (i.e., $\delta_i \rightarrow \delta$), which are ground states when $\delta/\Omega \gtrsim 1$. When $\delta / \Omega \lesssim 1$ the ground state transitions to a disordered phase dominated by Rabi oscillations, which will later become a focus in this work. 
In $1$D, ordered phases of $\ket{r}$ excitations crystallize to form a series of ground states labeled $\mathbb{Z}_q$ with commensurate (integer) spatial periods $q \in \{2, 3, 4, \ldots\}$ and rational fraction densities $\rho \sim \frac{1}{q}$ for increasing $R_b$, as shown in Fig.~\ref{TL behavior 1D and 2D}(c)~\cite{Bak1982,Rader2019}. In bulk systems, phase transitions between these crystalline phases are predicted to be separated by a gapless Luttinger liquid phase known as the floating phase~\cite{Rader2019}. It is characterized by incommensurate, but ordered, filling of $\ket{r}$, yielding continuous variation of $\rho$, correlation functions, and the spatial period of density fluctuations as a function of $R_b$. These quantities interpolate between the rational $\sim \frac{1}{q}$ values in the adjacent commensurate crystalline phases due to the gapless, continuous nature of the floating phase~\cite{Rader2019}. Fig.~\ref{TL behavior 1D and 2D}(d) shows the variation of spatial period of density fluctuations between the $\mathbb{Z}_3$ and $\mathbb{Z}_4$ phases. 

In contrast, experiments on finite $1$D lattices find that these essential properties of the floating phase are destroyed by finite-size boundary effects. The wavevector of the density fluctuations varies discretely as a function of $R_b$ and is strictly quantized to rational fractions of the system size, $k/(2\pi) \sim z/L$ for integers $z$ (see Fig.~\ref{TL behavior 1D and 2D}(d)). This is caused by strong pinning of $\ket{r}$ excitations at the edges due to their reduced interaction energy compared to atoms on the interior of the lattice~\cite{Zhang2025}. Such pinning collapses the continuous set of low-energy ``floating'' states into a small finite set of states in which $\ket{r}$ excitations do not get within $\sim R_b$ distance of the pinned edge excitation.

\begin{figure}[t!]
    \centering

    \includegraphics[
        width=0.75\linewidth,
        keepaspectratio
    ]{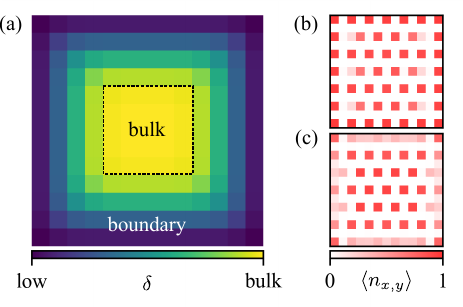}

    \vspace{1.1em}

    \includegraphics[
        width=\linewidth,
        keepaspectratio
    ]{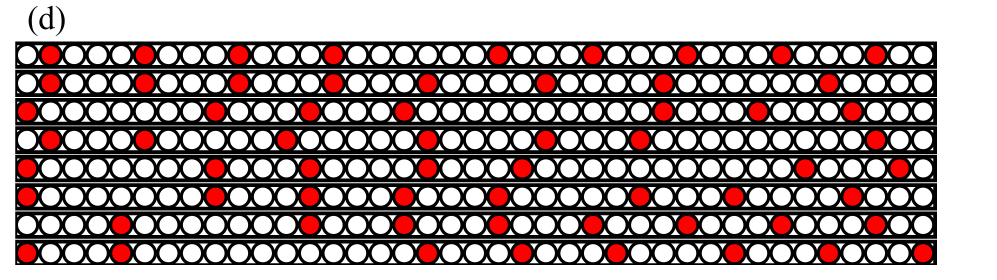}

    \vspace{1.1em}

    \includegraphics[
        width=\linewidth,
        keepaspectratio
    ]{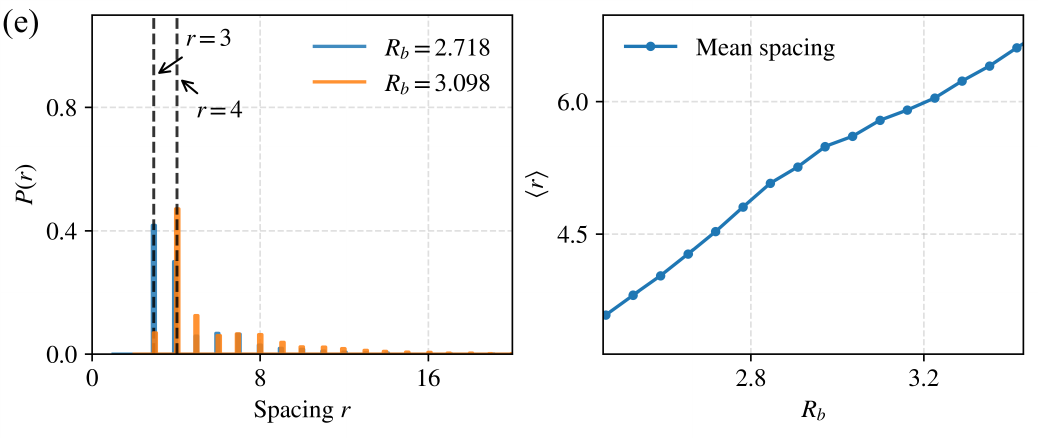}

    \caption{
        Disordered boundary subsystems. (a) Schematic illustration of the spatial variation of the detuning $\delta_i$ on a 2D square lattice. (b) and (c) compare the spatially resolved excitation density of a star phase ground state obtained with a uniform Hamiltonian (b) and non-uniform linear variation of $\delta_i$ (c). (d) Eight dominant classical configurations sampled from the $L=121$ disordered phase DMRG ground state at $(R_b,\delta) = (3.10, ~1.59)$, ordered vertically. Red (white) circles denote $\ket{r}$ ($\ket{g}$) states; $40$ sites within the center of the lattice are shown.
        (e) Distribution $P(r)$ of distance between nearest-neighbor excitations in the disordered phase ($R_b = 2.72, ~3.10$) together with the expected value of separation distance $\langle r \rangle$ as a function of $R_b$ (line cut at $\delta = 1.59$).
    }
    \label{disordered and floating phase}
\end{figure}

Numerical simulations of the $2$D square lattice predict that this system similarly supports a variety of ordered crystalline phases of $\ket{r}$ excitations. The situation is significantly complicated by the fact that the $2$D geometry admits crystalline orders with the same density but different symmetries, leading to competing low-energy states. The ground state phase diagram shown in  Fig. \ref{TL behavior 1D and 2D}(a) reports the stable orders in the thermodynamic limit, with a notably large region of stability for the $1/4$-density star phase (red).
However, in experiments on finite $2$D arrays of up to $\sim200$ atoms, the boundary effects are even more pronounced than in $1$D. The lower interaction energy at the edges encourages denser packing of $\ket{r}$ excitations around the boundary than is energetically favorable in the interior of the system (Fig.~\ref{disordered and floating phase}(b))~\cite{ORourke2023,Kalinowski2022}. This destabilizes the bulk ordered states at $R_b \geq 1.6$ in favor of competing low-energy orders~\cite{ORourke2023}. Fig. \ref{TL behavior 1D and 2D}(b) shows the ground state phase diagram of a $13 \times 13$ square lattice in a region dominated by the star phase in the bulk. On the finite lattice, a new $1/4$-density order with different symmetry, called the square phase, becomes the ground state over nearly all of the relevant $(\delta, R_b)$ parameter space.

These are just a few examples of the drastic differences that emerge between predicted bulk physics in $1$D and $2$D and their finite-array manifestations. This motivates the  development of an experimentally practical strategy to mitigate the boundary effects present in finite arrays. In this work, we propose a technique that addresses two key requirements: (i) elimination of pinned $\ket{r}$ excitations at the edges of the system, and (ii) a physically unbiased mechanism to drive the edge atoms into configurations that are compatible with the true bulk order. We will show that this can be achieved by
simulating a spatially non-uniform Hamiltonian using local control of the on-site detuning $\delta \rightarrow \delta_i$~\cite{Chen2023, Oliveria2025, Manovitz2025}. Specifically, a region in the center of the array can be chosen which retains the desired bulk Hamiltonian parameters, i.e., $\delta_i = \delta_{\mathrm{bulk}}$, while the boundary area is smoothly tuned into a parameter regime corresponding to the disordered phase via decreasing $\delta_i$. A schematic example of the spatial profile of $\delta_i$ is shown for a $2$D square lattice in Fig~\ref{disordered and floating phase}(a). 

By using a sufficiently smooth variation of $\delta_i$, sharp energetic ``interfaces'' are removed from the Hamiltonian. We will show that this prevents $\ket{r}$ excitations from becoming pinned at specific lattice sites, instead allowing all the atoms to fluctuate between $\ket{g}$ and $\ket{r}$ as they would in a bulk system. Additionally, we will show that such a simple strategy is useful because it does not require fine tuning and in fact is agnostic to the structure of the bulk ground state, allowing it to be applied in a general context when the true ground state is not already known, e.g., from computation. This result relies on the detailed microscopic properties of the disordered phase. 

Typically the disordered phase is understood as a direct analog to the gapped paramagnetic phase of the transverse field Ising model, with each atom (or, spin) pointing along the direction of the Rabi term (or, field)~\cite{Weimer2010}. In the Rydberg array picture, the state appears featureless when analyzing the expected local density $\langle \hat{n}_i\rangle$, spatial structure of $\ket{r}$ excitations, and structure factors~\cite{Rader2019,ORourke2023}. While the analogy to the paramagnetic phase is valid in the regimes $\delta < 0$ and $\delta \ll \Omega$, Figs.~\ref{TL behavior 1D and 2D}(a) and (c) clearly reveal that it remains stable for $\delta$ values equal to and exceeding $\Omega$.
In this regime, the disordered phase has a correlated ground state~\cite{ORourke2023} consisting of a significant superposition of low-energy configurations (i.e., in the $\{\ket{g}, \ket{r} \}$ basis). 
To characterize this structure, we analyze the matrix product state (MPS) representation of the ground state obtained from DMRG for representative parameters $(R_b, \; \delta, \; \Omega) \equiv (3.10,\; 1.59,\; 1)$ on a $L=121$ $1$D lattice with a uniform Hamiltonian (i.e., $\delta_i \rightarrow \delta$). Using the perfect sampling algorithm for MPS~\cite{Stoudenmire2010,Ferris2012}, we find that the wavefunction is composed of a large set of unique configurations with nearly equal probabilities, which is consistent with recent experimental measurements~\cite{Zhang2025}. Representative configurations with the largest probabilities are shown in Fig.~\ref{disordered and floating phase}(d). Although the ground state still lacks long-range order and appears featureless to local measurements in the $\delta \approx \Omega$ regime, the sampled ensemble contains clear structure. Ordered short-range clusters are prevalent, with $\ket{r}$-spacing set by the value of $R_b$, while domain walls of $\ket{g}$ sites break up the long-range order. For example, in Fig~\ref{disordered and floating phase}(d) significant ordered clusters appear with an $\ket{r}$-spacing of $4$ because the value of $R_b=3.1$ is proximal to the $\mathbb{Z}_4$ ordered phase. The nature of the disordered ground state in the $\delta \approx \Omega$ regime can therefore be understood as a broad superposition of configurations which contain local imprints of proximal ordered phases but also sufficient density of $\ket{g}$-domain walls to enable their mixing into a superposition. The central insight of this work is that such a state will have sufficient \textit{local} overlap with any proximal ordered phase that strong interactions between an ordered ``bulk'' region and a disordered ``boundary'' region will select out the most favorable subset of quasi-ordered configurations from the pure disordered state.

To further emphasize the flexibility of the dominant configurations making up the disordered superposition, in Fig. \ref{disordered and floating phase}(e) we report the distribution of nearest-neighbor excitation spacings (hereby denoted as $P(r)$) extracted from the sampled classical configurations.
For $R_b$ values near the $\mathbb{Z}_3$ and $\mathbb{Z}_4$ phases, the spacing distribution is peaked at $r = 3$ and $r = 4$, respectively, while also containing a long tail at larger $r$ values. This indicates that the disordered phase flexibly contains strong local signatures of the nearby commensurate orders in addition to a proliferation of domain walls. Moreover, by examining the mean spacing $\langle r \rangle$ across a range of $R_b$ values, we can see that the dominant configurations smoothly deform as a function of $R_b$. This smoothness highlights the continuous nature of how the disordered superposition responds to changes in the interactions, which is the key property enabling it to serve as an unbiased and responsive boundary subsystem interacting with an ordered bulk subsystem.

A straightforward demonstration of this behavior is shown for a $13 \times 13$ $2$D square lattice in Fig.~\ref{disordered and floating phase}(b)-(c). In (b), the real-space excitation density is plotted for the ground state of a spatially uniform Hamiltonian with parameters that stabilize the star phase $(R_b, \delta) = (1.9, 4.0)$. The boundary has a strongly pinned configuration of excitations that is incommensurate with the bulk star order, leading to boundary-bulk frustration~\cite{Kalinowski2022}. In (c), the same data is shown for the ground state of a non-uniform Hamiltonian containing a simple linear variation of $\delta_i$ between $\delta_i=4.0$ in the bulk region to $\delta_i = 1.8$ at the boundary, as shown schematically in Fig.~\ref{disordered and floating phase}(a). Here, the interaction between the bulk subsystem and the disordered boundary subsystem causes the boundary atoms to adopt a small superposition commensurate with the bulk order, allowing for a cleaner stabilization of the star order devoid of frustration in the bulk.

\begin{figure}[t!]
    \centering

    \includegraphics[width=1.05\linewidth, keepaspectratio]
    {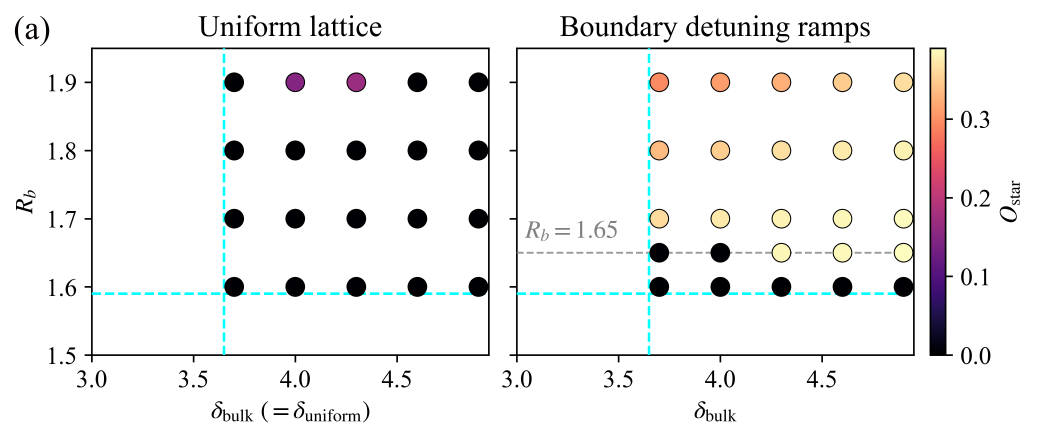}
    
    \vspace{0.8em}
    
    \includegraphics[width=\linewidth, keepaspectratio]
    {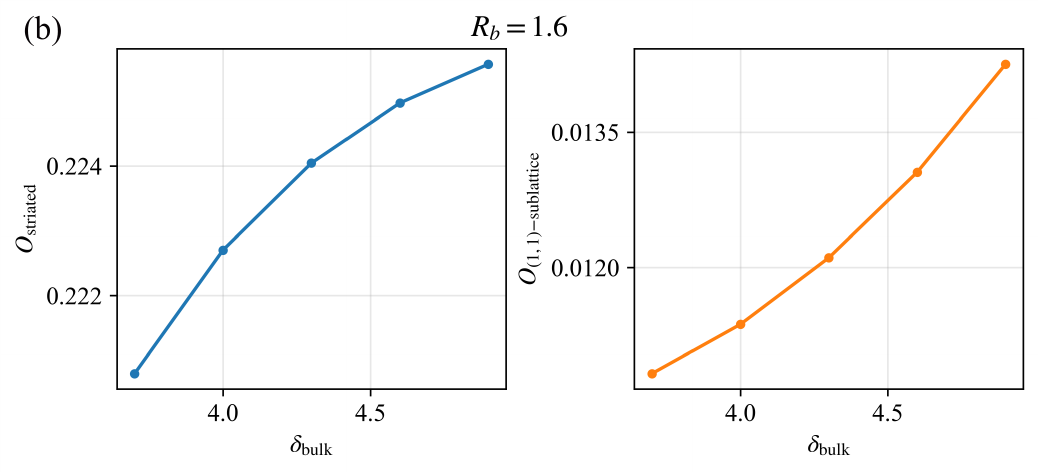}

    \caption{
        (a) Comparison of star ground state order parameter on a $13 \times 13$ array for a Hamiltonian with spatially uniform $\delta$ and one with disordered boundary subsystems. The non-uniform $\delta_i$ uses $n_{\mathrm{boundary}} = 4$ outer rings, $\delta_{\mathrm{boundary}} = 1.8$, and interface detuning mismatch $\alpha = 0.05$. Dashed cyan lines denote the numerically computed region. (b) Striated and $(1,1)$-sublattice order parameters as a function of bulk detuning $\delta_{\mathrm{bulk}}$ at $R_b = 1.6$.
    }
    \label{2D results}
\end{figure}

As discussed earlier, when considering a typical spatially uniform Hamiltonian the ground state phase diagram of the $2$D square lattice is severely affected by the behavior of the boundary.
In the region $1.6 \leq R_b \leq 1.9$ and $3.7 \leq \delta_{\mathrm{bulk}} \leq 4.9$, the star phase is the stable ground state in the thermodynamic limit. However, the strong interactions between the bulk and the densely packed, pinned excitations at the boundary of finite systems broadly favor the stability of the square phase ground state which has a structure commensurate to the boundary (Fig.~\ref{TL behavior 1D and 2D}). In Fig.~\ref{2D results} we demonstrate that by introducing the disordered boundary subsystems, we can recover the correct thermodynamic limit order by stabilizing the star ground state across the entire parameter regime.

In Fig. \ref{2D results}(a) we compare the ground state phase diagram obtained on a $13\times 13$ square lattice for a spatially uniform Hamiltonian to a non-uniform Hamiltonian employing a central $5 \times 5$ bulk region with $4$ surrounding rings of boundary sites. The detuning profile is a simple linear variation of $\delta_i$ between $\delta_{\mathrm{bulk}}$ in the center to $\delta_{\mathrm{boundary}}=1.8$ at the boundary, with a mismatch $\alpha=0.05$ at the interior bulk-boundary interface. Note that the $5\times 5$ bulk subsystem geometry does not artificially favor the star phase over square since it is perfectly commensurate with the finite square order.
To identify the star phase, we employ the order parameter (OP) defined in Ref. \cite{ORourke2023}:
\begin{equation}
    O_{\mathrm{star}} := \displaystyle\sum_{x, y} (\langle \hat{n}_{x, y} \rangle - \langle \hat{n}_{y, x} \rangle)^2/N_{\mathrm{bulk}}, 
\end{equation}
where $N_{\mathrm{bulk}}$ denotes the size of the bulk region. We find that the disordered boundaries yield sharp stability of the star-ordered ground state across nearly the entire region $1.65 \leq R_b \leq 1.9$ and $3.7 \leq \delta_{\mathrm{bulk}} \leq 4.9$. Comparing the magnitude of the OP with the uniform Hamiltonian reveals the robustness of the star order in the presence of the boundary subsystems, as compared to the weak, frustrated order in the uniform case.

Our tests with smaller values of $\delta_{\mathrm{boundary}}$ reinforced the greater importance of a gradual detuning profile than a very small value of $\delta_{\mathrm{boundary}}$. Although having both properties is ideal, the current size of state of the art experiments makes this impossible in $2$D, where the length scale between the edge of the bulk subsystem and the boundary scales as the square root of the number of atoms in the boundary subsystem. The gradual variation of the detuning profile prevents local mismatch of energy scales between neighboring lattice sites, which inhibits the pinning of excitations like we see along the edge of the uniform case (Fig.~\ref{disordered and floating phase}(b)).

We additionally scrutinize the results of the boundary subsystem phase diagram at $R_b = 1.6$, where the star phase is stable in the thermodynamic limit but not in our simulations. A close inspection reveals features consistent with the striated phase, which shares the dominant ordering of the square phase but, crucially, also requires weak excitation-density fluctuations on the $(1, 1)$-sublattice along rows where excitations are otherwise suppressed in the square phase (Fig.~\ref{TL behavior 1D and 2D}(a))~\cite{Samajdar2020, ORourke2023}. Fig.~\ref{2D results}(b) reports a large, non-zero value for an OP that detects the presence of the square and/or striated order, $O_{\mathrm{striated}}$, defined in Ref.~\cite{Samajdar2020} and the SM. To distinguish between the two, Fig.~\ref{2D results}(b) also reports the average density of excitations on the $(1,1)$-sublattice (denoted as $O_{(1, 1)-\mathrm{sublattice}}$), which are $0$ in the square phase~\cite{ORourke2023}. While $O_{\mathrm{(1, 1)-\mathrm{sublattice}}}$ remains small at $R_b = 1.6$, it is consistently nonzero (and monotonically increases) across various $\delta_{\mathrm{bulk}}$ values, indicating the presence of weak but systematic density fluctuations on the $(1, 1)$-sublattice. Additional numerical tests in the SM demonstrate that these small fluctuations are not negligible; if suppressed to exactly $0$, the star phase becomes the ground state at $R_b=1.6$. In the thermodynamic limit, only the striated and star phases are stable while the square phase is not. The boundary subsystem simulations recover this important feature of the thermodynamic limit phase diagram, while introducing only a minor shift in the phase boundary between the striated and star phases.

To test the efficacy of the boundary subsystem protocol in a more demanding setting, we now turn to the problem of recovering the floating phase ground state on finite $1$D lattices. In this case, the thermodynamic-limit ground state is a gapless, correlated superposition of configurations as opposed to the preceding example of the $2$D star phase which is a mean-field state.
In finite systems with a uniform Hamiltonian, boundary effects strictly discretize the allowed density-wave fluctuations in the floating phase~\cite{Zhang2025}, restricting access to the continuous manifold of incommensurate floating-phase states (Fig.~\ref{TL behavior 1D and 2D}(d)). This can be understood in terms of pinned excitations at the boundary strongly interacting with the bulk, allowing only density-wave orders with excitations spaced by a distance of $\sim R_b$ from the edge to retain a low energy.

\begin{figure}[t!]
    \centering

    \includegraphics[width=\linewidth, keepaspectratio]
    {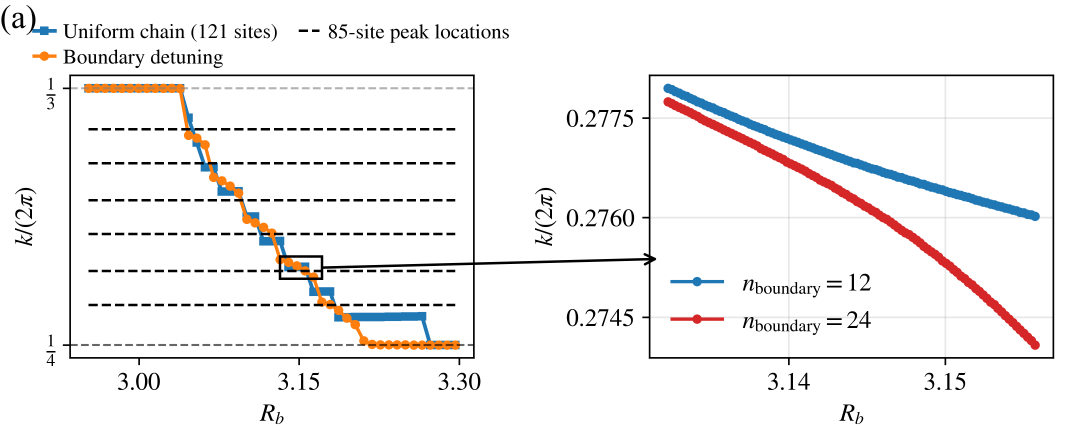}

    \includegraphics[width=\linewidth, keepaspectratio]
    {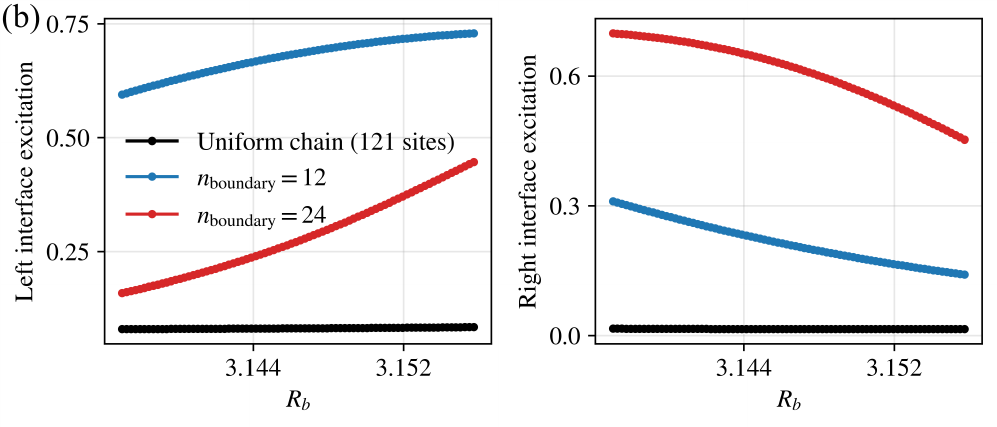}


    \caption{
        (a) Density-wave wavevector $k$ vs.~$R_b$ for an $L = 121$ chain for both uniform and non-uniform $\delta_i$ profiles ($n_{\mathrm{boundary}} = 24$). Peak locations for the uniform $L = 85$ chain are shown as black dashed lines for comparison. The highlighted rectangular region indicates a region with continuous variation. The right panel shows a high resolution $R_b$ scan within this highlighted region for multiple boundary subsystem sizes ($\delta_{\mathrm{boundary}} = 0.2$). (b) Excitation density $\langle n_i \rangle$ vs.~$R_b$ for the atoms at the interface between the bulk and boundary subsystems. The uniform chain data shows $\langle n_i \rangle$ for the $24$th site from either edge of the finite chain.  
    }
    \label{1D results}
\end{figure}

To relax this constraint, in Fig.~\ref{1D results} we study a $1$D lattice with fixed length $L=121$, holding the bulk region at $\delta = 4.06$ and imposing a linear variation of $\delta_i$ over boundary subsystems of size $n_{\mathrm{boundary}} = 12$ or $24$ on either side of the bulk region. The minimum value of $\delta_i$ at the edge is $\delta_{\mathrm{boundary}} = 0.2$, with a small bulk-boundary interface mismatch $\alpha = 0.1$. We characterize the computed ground states in the region $2.9 < R_b < 3.3$ by extracting the wavevector $k$ of the density-wave order in Fig.~\ref{1D results}(a). Within the regions corresponding to the $\mathbb{Z}_3$ and $\mathbb{Z}_4$ phases $k/(2\pi)$ takes rational values of $1/3$ and $1/4$, respectively. In the intermediate region corresponding to the floating phase, the uniform Hamiltonian ground states display the expected series of sharp, discrete $k$-plateaus. In contrast, the ground states in the presence of disordered boundary subsystems display piecewise-continuous variation of $k$ with $R_b$, with each point corresponding to a physically distinct bulk mode. Zooming in on one of the quasi-continuous regions with a fine-$R_b$ scan reveals that the variation of $k$ is truly continuous, suggesting a type of gapless behavior.

Comparing the results from different boundary region sizes in Fig.~\ref{1D results}(a), we see that the larger boundary region permits bulk ground state orders with a larger continuous range of $k$ values. This is corroborated in Fig.~\ref{1D results}(b), which shows that the atom at the bulk-boundary interface adopts a wider range of possible excitation densities $\langle n_i\rangle$ when the boundary subsystem is larger. This demonstrates how the disordered boundary subsystem is able to provide an unbiased, flexible set of boundary conditions to the bulk subsystem, governed only by the bulk order and the strong interactions between the boundary and bulk subsystems.
Additionally, it highlights the importance of gradual variation of $\delta_i$ within the boundary subsystem. Comparing to the same atom in the ground states of the uniform Hamiltonian, its excitation density is completely static to variations in $R_b$ due to the pinned order. 

To place these finite lattice results in context, Fig. \ref{TL behavior 1D and 2D}(d) shows the wavevector trajectory for a uniform $1009$ atom lattice, which exhibits an almost perfectly continuous $k(R_b)$ evolution across the floating phase (on the scale of these plots). Remarkably, by using sufficiently large disordered boundary subsystems, the wavevectors of ground states on the $121$-site lattice begin to closely track this thermodynamic-limit behavior by enabling more faithful access to the full manifold incommensurate density-wave orders.
This demonstrates that modest, experimentally feasible boundary modifications can recover floating-phase physics on lattice an order of magnitude smaller than would otherwise be required (see SM for an extended set of numerical results and an algebraic treatment of the boundary conditions.)

In conclusion, we have shown that the disordered phase can be used to great effect as a simple boundary subsystem that substantially mitigates the typical finite-size effects observed in current Rydberg atom experiments. By simulating the ground states of a non-uniform Hamiltonian with a generic linear variation of $\delta_i$ in the boundary subsystem, the bulk subsystem can substantially recover many aspects of the physics present in the thermodynamic limit. In $2$D, we demonstrated that the stability of the star phase ground state can be recovered, while in $1$D the sharp discretization of density-wave ordering in the floating phase region can be made quasi-continuous. This is made possible by the intrinsic structure of the disordered phase in the $\delta \approx \Omega$ regime, which naturally contains a broad superposition of configurations containing significant local overlap with proximal ordered phases. While the results here focus on ground states, we expect that this idea will open many new avenues of investigations, including  dynamical phenomena, topological order in $2$D, and designing more sophisticated protocols to control the boundary subsystem.

\begin{acknowledgments}
This research was supported by startup funding from Brown University. Computations were performed using resources at the Center for Computation and Visualization, Brown University. 
\end{acknowledgments}

\bibliography{References}

%

\end{document}


\title{Supplemental Material for ``Stabilization of bulk quantum orders in finite Rydberg atom arrays''}

\author{Yash M. Lokare}
\affiliation{Department of Physics, Brown University, Providence, Rhode Island 02912, USA
}%
\author{Matthew J. Coley-O'Rourke}%
\affiliation{Department of Chemistry, Brown University, Providence, Rhode Island 02912, USA
}%

\date{\today}
\maketitle

\tableofcontents


\section{\label{Computational methods} Computational details}

\begin{figure*}[h]
    \centering
    \includegraphics[width=0.48\linewidth, keepaspectratio]
    {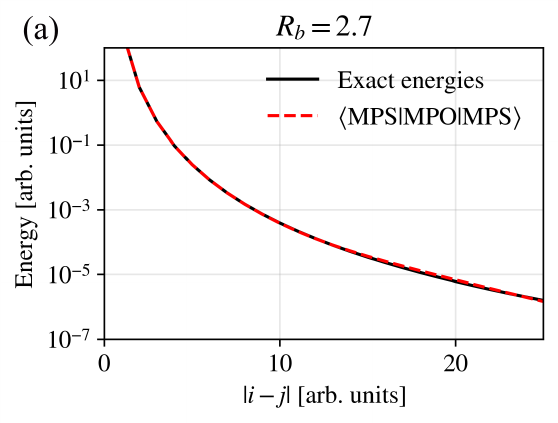}
     \includegraphics[width=0.48\linewidth, keepaspectratio]
    {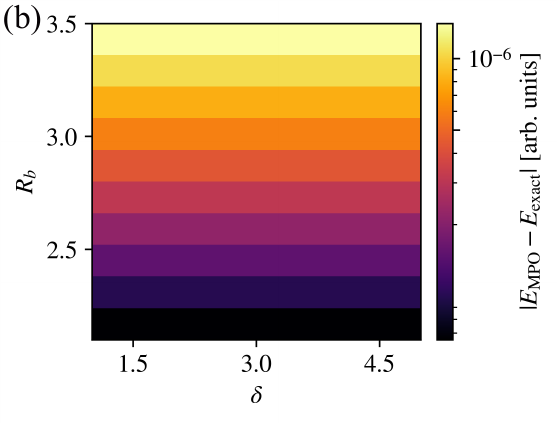}
    \caption{
        Validation of the MPO representation of $1/r^6$ interactions using a $24$-term exponential decomposition for a $121$-site chain. \textbf{(a)} Exact interaction energies vs. $\bra{\mathrm{MPS}} \mathrm{MPO} \ket{\mathrm{MPS}}$ for states with two excitations separated by $r = \{1, 2, \ldots, N - 1\}$. We set $R_b = 2.7$. \textbf{(b)} Total energy comparison for full MPO evaluations on a configuration with excitations placed every five sites. 
    }
    \label{benchmark MPO accuracy}
\end{figure*}

The phase diagram in Fig 1(a) of the main text, depicting the reference thermodynamic limit ground states in $2$D, was computed using the $\Gamma$-point DMRG method detailed in Ref.~\cite{ORourke2023}. All DMRG simulation details are kept identical for consistency. For all other results in the main text we employ the finite-system DMRG algorithm \cite{White1992, Schollwock2005} to find the ground state. For $L=85, 121$ lattices in $1$D, we use an in-house implementation of finite-system DMRG~\cite{LokareGithub} that employs the standard exponential fitting technique to represent the long-range interactions as a matrix product operator (MPO)~\cite{Crosswhite2008}. We use $24$ exponentials for a high-accuracy fit, as shown in Fig.~\ref{benchmark MPO accuracy}. For $L=1009$ lattices in $1$D and all $2$D arrays, we use the {\tt BLOCK2} DMRG simulation package~\cite{Zhai2023} and build the long-range MPO using a more sophisticated SVD-based method for efficiency~\cite{Chan2016}. The $2$D lattice is mapped onto a $1$D MPS using a standard ``snake'' ordering. In all simulations other than Fig 1(a), we use open boundary conditions (OBCs) to match experimental conditions and long-range Rydberg interactions are retained without truncation in both $1$D and $2$D. 

For $1$D simulations, lattice sizes $L = 85, 121$, and $1009$ are chosen to be commensurate with the $\mathbb{Z}_3$ and $\mathbb{Z}_4$ ordered phases so as to accurately probe the intervening floating phase. When studying the non-uniform Hamiltonians, the boundary subsystem sizes $n_{\mathrm{boundary}}$ are selected such that the resulting bulk region $N_{\mathrm{bulk}}$ is likewise compatible with these commensurate ordered phases. All $1$D calculations employ a bond dimension of $\chi = 150$.

For the $2$D simulations, we focus on a finite $13 \times 13$ square lattice as an experimentally relevant and computationally tractable geometry. The boundary subsystems are of width $n_{\mathrm{boundary}} = 4$ sites on all sides, resulting in a central $5 \times 5$ bulk region. This bulk size is chosen to be compatible with both the square ($2 \times 2$), star ($4 \times 2$) and striated ($2 \times 2$) ordering patterns, enabling a fair comparison of competing crystalline phases. We employ a bond dimension of $\chi = 500-600$ and typically find that we must perform $\sim 60$ finite-system DMRG sweeps to achieve well-converged ground-state energies and wavefunctions. To reduce the likelihood of convergence to metastable local minima, we follow the typical protocol to introduce a small amount of noise into the MPS tensors during the initial sweeps.

\section{Computing wavevectors of incommensurate density-wave orders}

Considering a $1$D lattice of $L$ sites, the density-wave ordering of the ground state can be described by the set of all the site-resolved excitation densities $\langle n_j \rangle$:
\begin{equation}
    \langle n_j \rangle \in \mathbb{R}; ~j = \{0, 1, \ldots, L - 1\}. 
\end{equation}
Taken together, we call this the ``density profile.'' Dominant wavevectors of the density profile can be computed by the Fourier transform (FT) of this discrete function $\langle n_j \rangle \equiv n[j]$. The standard discrete FT evaluates the structure factor only at the discrete Fourier modes allowed by the lattice: 
\begin{equation}
    \tilde{n}[m] = \displaystyle\sum_{j = 0}^{L - 1} n[j] ~e^{-2\pi m j/L}, 
\end{equation}
where $m \in \{0, 1, \ldots, L - 1\}$ and the corresponding wavenumbers $k_m = 2\pi m/L$. The associated structure factor is $S[m] = |\tilde{n}[m]|$. While this is sufficient for global spectral features, it limits the precision of peak localization to the spacing $\Delta k = 2\pi/L$. To achieve sub-bin resolution and precisely determine the incommensurate wavevectors, we compute a continuous/oversampled FT by evaluating the discrete FT at a dense grid of wavenumbers:
\begin{equation}
    \tilde{n}(k) = \displaystyle\sum_{j = 0}^{L - 1} n[j] ~e^{-ik j},
\end{equation}
where $k \in [0, 2\pi)$. We define a dense wavenumber grid with $M = \omega L$ points, where $\omega$ is the oversampling factor (i.e., $\omega$ determines how fine the grid is). We have
\begin{equation}
    k_{\ell} = \frac{2 \pi \ell}{\omega L}, 
\end{equation}
where $\ell \in \{0, 1, \ldots, \omega L - 1\}$. The quasi-continuous structure factor is then evaluated as
\begin{equation}
    S(k_{\ell}) = |\tilde{n}(k_{\ell})| = \Bigg|\displaystyle\sum_{j = 0}^{L - 1} n_j ~e^{-i k_{\ell} j}\Bigg|. 
\end{equation}

In practice, we know \textit{a priori} that ground states in the floating phase will have a single dominant density wavevector~\cite{Rader2019}, and our goal is to precisely compute its value. In this way, we are not trying to extract more information from $n[j]$ than is fundamentally allowed. This concept is regularly employed in signal processing via the so-called ``discrete time Fourier transform.''

Specifically, the procedure we apply is as follows: 
First we compute the dominant discrete mode $k_n^*$ from the standard discrete FT. Then we construct a local high-resolution window around it, 
\begin{equation}
    k \in [k_n^* - \Delta k, ~k_n^* + \Delta k], 
\end{equation}
sampled with a fine mesh (set by $\omega$). Within that window, we compute $S(k)$ quasi-continuously to locate the true maximum
\begin{equation}
    k_{\mathrm{peak}}^{\mathrm{refined}} = \arg \max\limits_{k \in [k_n^* - \Delta k, ~k_n^* + \Delta k]} S(k). 
\end{equation}
This gives the refined peak location. To resolve the continuous variation of the peak location in main text Fig. 4(a), we set $\omega = 1000$.

\section{Characterization of the disordered phase}

To characterize the structure of the disordered ground state in the $\delta \approx \Omega$ regime, we analyze the distribution of classical $\{ \ket{g_i}, \ket{r_i}\}$ configurations sampled from the variationally optimized MPS using the perfect sampling algorithm for MPS~\cite{Stoudenmire2010,Ferris2012}. For a normalized ground-state MPS $\ket{\psi_{\mathrm{g}}}$ defined on a lattice of $L$ sites, the sampling procedure generates computational-basis configurations with exact probabilities encoded in the MPS wavefunction.

To characterize the nature of the disordered phase, we generate an ensemble of $\mathcal{O}(10^4)$ of the most probable classical configurations using the sampling algorithm. We can then approximately analyze the spacing between excitations in the ground state MPS directly from the ensemble. This approach yields a transparent probabilistic characterization of the spacing structure. Each sampled configuration can be represented as a binary string
\begin{equation}
    s^{(k)} = (s_1^{(k)}, s_2^{(k)}, \ldots, s_L^{(k)}); ~s_j^{(k)} = \{0, 1\}, 
\end{equation}
where $s_{j}^{(k)} = 1$ denotes a Rydberg excitation. The sampling procedure draws each configuration with its exact Born probability
\begin{equation}
    p(s^{(k)}) = |\psi_{\mathrm{g}}(s^{(k)})|^2, 
\end{equation}
evaluated on-the-fly as part of the sampling process. For a given configuration $s^{(k)}$, we identify the ordered set of positions of excited sites and compute all nearest-neighbor separations: 
\begin{equation}
    \mathcal{R}(s^{(k)}) = \{r_1^{(k)}, r_2^{(k)}, \ldots, r_{N-1}^{(k)}\},
\end{equation}
where $N$ is the number of excitations in the sample $s^{(k)}$. Each spacing value reflects a specific physical separation within that configuration. To build the global spacing distribution, we aggregate spacing values from all configurations, weighting each contribution by the probability of observing the configuration itself. Mathematically, the probability that two adjacent excitations are separated by distance $r$ is
\begin{equation}
    P(r) = \frac{1}{\mathcal{N}} \displaystyle\sum_{k} p(s^{(k)}) \displaystyle\sum_{\ell} \delta(r - r_{\ell}^{(k)}), 
\end{equation}
where the inner sum runs over all spacings $\{r_{\ell}^{(k)}\}$ in configuration $s^{(k)}$. The normalization constant 
\begin{equation}
\mathcal{N} = \displaystyle\sum_{r} \biggl[\displaystyle\sum_{k} p(s^{(k)}) \displaystyle\sum_{\ell} \delta(r - r_{\ell}^{(k)})\biggr]
\end{equation}
enforces $\displaystyle\sum_{r} P(r) = 1$. The average spacing between excitations is then obtained as the expectation value of $r$ under this distribution:
\begin{equation}
    \langle r \rangle = \displaystyle\sum_{r} r \cdot P(r), 
\end{equation}
providing a direct measure of the dominant excitation length scale in the quantum ground state. 

\section{Algebraic analysis of boundary conditions in the floating phase}
\label{discretization floating phase}

In a density-ordered phase, the local excitation density profile can be modeled by the general form~\cite{Zhang2025, Garcia2025, Liao2025},
\begin{equation}
    n[j] = n_0 + A \cos(kj + \Phi), 
\end{equation}
where $n_0$, $A$, $k$, and $\Phi$ are set by the microscopic details and boundary conditions. Let's consider the boundary conditions at the two ends of the bulk region. The angles $(kj + \Phi) \equiv \theta$ assume the form,
\begin{eqnarray}
    \theta_1 = k \cdot 1 + \Phi, \\
    \theta_2 = k \cdot N_{\mathrm{bulk}} + \Phi, 
\end{eqnarray}
where the end  sites correspond to $j = 1, ~N_{\mathrm{bulk}}$. The behavior of the excitations at the edge of the system will determine the values of $k$ and $\Phi$. In general, cosine takes values $[-1,1]$, so we have,
\begin{equation}
    \cos(\theta_1) = C_1, ~\cos(\theta_2) = C_2,  
\end{equation}
where $C_1, C_2 \in [-1, 1]$ are values given by the state of the edge atoms. Solving for $\theta$ modulo $2\pi$, we obtain
\begin{equation}
    \theta_1 = \arccos(C_1) + 2\pi n_1, ~~\theta_2 = \arccos(C_2) + 2\pi n_2,
\end{equation}
where $n_1, n_2 \in \mathbb{Z}$. Subtracting the two equations, we obtain
\begin{equation}
    k \cdot (N_{\mathrm{bulk}} - 1) = \pm \arccos(C_2) \mp \arccos(C_1) + 2\pi m,
    \label{eq:general_solution_boundary}
\end{equation}
where $m = n_2 - n_1 \in \mathbb{Z}$. 

\subsection{Pinned boundaries}
In the case of a uniform Hamiltonian, the edge atoms are pinned in the excited state. Given $A, n_0>0$, we can describe the pinned boundaries by fixing $C_1 = C_2 = 1$. In this case, Eq.~\eqref{eq:general_solution_boundary} reduces to
\begin{equation}
    k = \frac{2 \pi m}{N_{\mathrm{bulk}}-1}.
\end{equation}
Since $m$ is an integer, this reveals the strict quantization of the allowed wavevectors $k$ on a finite $1$D lattice.

\subsection{Flexible boundaries}
In the case of a non-uniform Hamiltonian, we have argued in the main text that the boundaries can be made ``flexible,'' or in other words, responsive to the bulk order. To describe this, the state of the edge atoms must be $R_b$-dependent. Defining the function $\Lambda(R_b) = \pm \arccos(C_2) \mp \arccos(C_1)$,  Eq.~\eqref{eq:general_solution_boundary} yields,
\begin{equation}
    \label{peak positions}
    k_{m, \Lambda} = \frac{2\pi m + \Lambda(R_b)}{N_{\mathrm{bulk}} - 1},
\end{equation}
Equivalently, we have
\begin{equation}
    k_{m, \Lambda} = \frac{2\pi}{N_{\mathrm{bulk}} - 1} \biggl(m + \frac{\Lambda(R_b)}{2\pi}\biggr). 
\end{equation}
In principle, $\Lambda(R_b)$ can take continuous values in the range $[0,2\pi)$ depending on the variation of $C_1, C_2$ with $R_b$. This precisely interpolates between adjacent integer values $m$ and $m+1$, allowing for a continuous spectrum of $k$-modes. If the edge atoms are only responsive to variations of $R_b$ within a restricted range of possible configurations, i.e., $C_1, C_2 \in [1, \epsilon]$, then the range of $k$ values will not be completely continuous but rather piecewise continuous. This is what we observe in main text Fig. 4(a).

\section{Extended numerical results}

\subsection{1D results}
\subsubsection{Stability of the floating phase in the non-uniform Hamiltonian}

\begin{figure*}[htpb!]
\includegraphics[width = 0.8\textwidth]{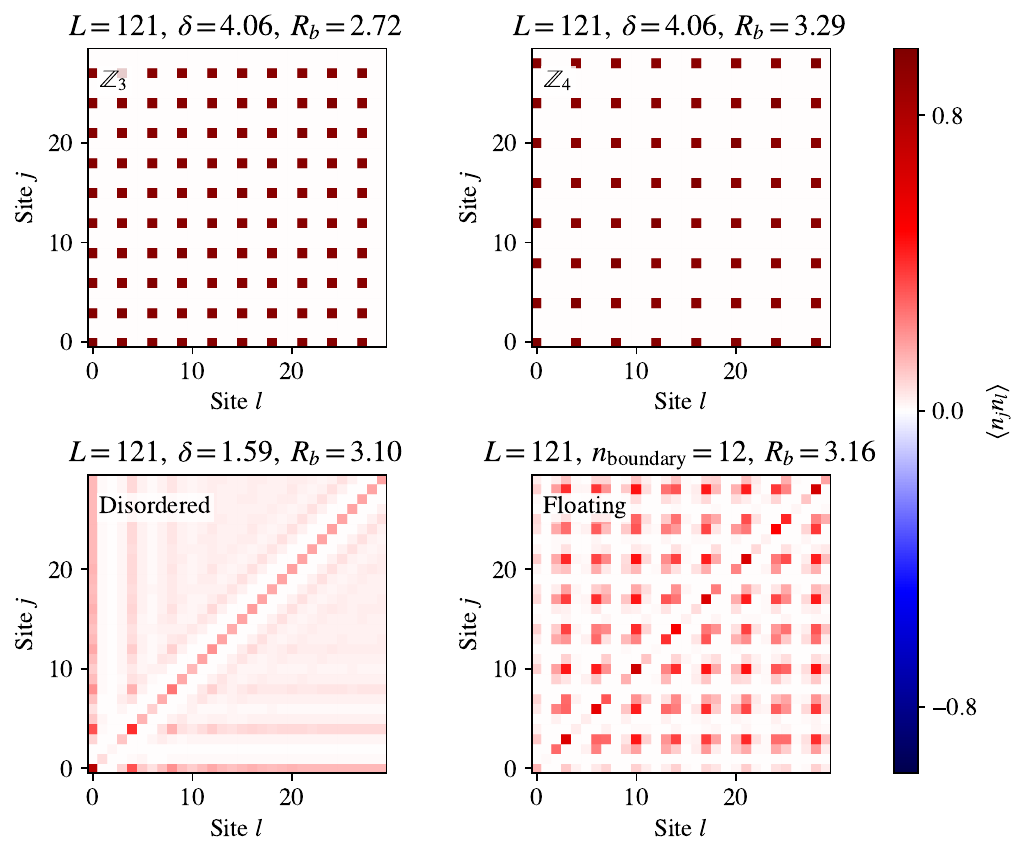}
\caption{Spatial correlation maps of the bulk region for $L = 121$ at representative points in the disordered, $\mathbb{Z}_3$, $\mathbb{Z}_4$, and floating phases. For the disordered, $\mathbb{Z}_3$, and $\mathbb{Z}_4$ phases, data from a uniform chain is shown. For the floating phase, results are obtained from a non-uniform chain with $\delta_{\mathrm{bulk}} = 4.06$, $\delta_{\mathrm{boundary}} = 0.2$, $n_{\mathrm{boundary}} = 12$, and $\alpha = 0.1$ interface detuning mismatch. In all cases, data for the $30$ sites in the bulk region is shown.}
\label{spatial correlation maps}
\end{figure*}

Boundary perturbations in $1$D Rydberg chains can, in principle, destabilize subtle quantum phases. Despite this sensitivity, we find that the floating phase remains remarkably robust even when the Hamiltonian is made explicitly non-uniform through variations in the local detuning $\delta_i$.
We examine the spatial structure of correlations using the site-resolved two-point correlation function $\langle n_j n_l \rangle$ ($1 \leq j, ~l \leq N_{\mathrm{bulk}}$). The corresponding spatial correlation maps for a chain of size $L = 121$, shown for representative points in the disordered, $\mathbb{Z}_3$, $\mathbb{Z}_4$, and floating phases in Fig. \ref{spatial correlation maps}, exhibit clear and qualitatively distinct behaviors across these phases. 

In the $\mathbb{Z}_3$ and $\mathbb{Z}_4$ phases, the correlation maps display perfectly ordered patterns with well-defined repeating unit cells corresponding to period-$3$ and period-$4$ density-wave order, respectively. These commensurate structures appear as rigid, phase-locked oscillations that are fixed to the lattice. In contrast, the disordered phase shows rapidly decaying correlations, with no persistent structure beyond short distances. 

The floating phase, however, is qualitatively distinct. The correlation map exhibits robust long-range order without a simple unit cell. The absence of simple periodic repetition indicates that the underlying density wave is incommensurate with the lattice spacing. Crucially, these incommensurate features are observed in the non-uniform chain ($n_{\mathrm{boundary}} = 12$). The absence of commensurate locking or short-range decay demonstrates that the boundary detuning variations do \textit{not} drive the system into the $\mathbb{Z}_3$, $\mathbb{Z}_4$, or disordered phases. 

\begin{figure}[htbp!]
    \centering



    \includegraphics[width=0.75\linewidth, keepaspectratio]
    {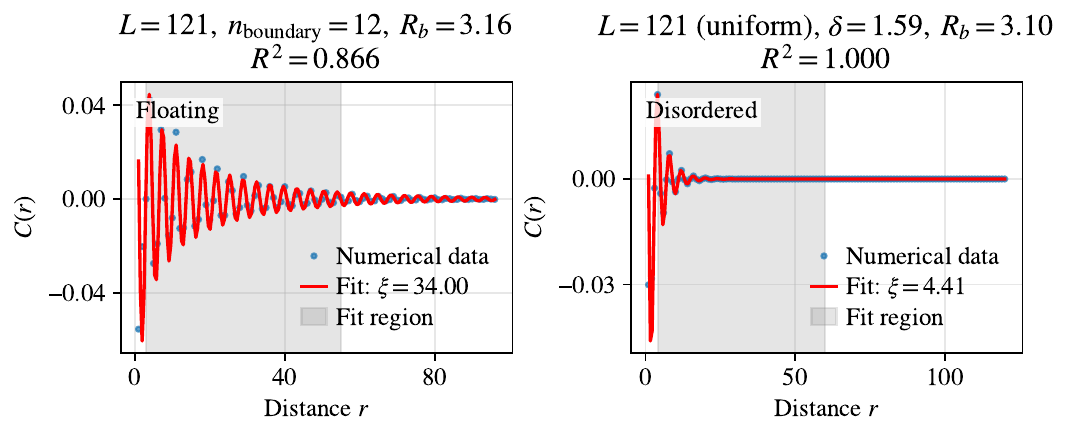}

    \caption{
        Ornstein–Zernike fits to the connected density–density correlator for $L = 121$ at representative points in the floating and disordered phases for a non-uniform and uniform chain, respectively. Parameters in the non-uniform setting: $\delta_{\mathrm{bulk}} = 4.06$, $\delta_{\mathrm{boundary}} = 0.2$, $n_{\mathrm{boundary}} = 12$, $\alpha = 0.1$ interface detuning mismatch; uniform-chain setting: $\delta = 1.59$. 
        The fit window spans $\sim 60\%$ of the bulk region.
    }
    \label{floating phase stability}
\end{figure}

To quantify the range of correlations in the disordered and floating phases, we extract correlation lengths by fitting the connected density-density correlator, 
\begin{equation}
    C(r) = \langle n_j n_{j + r} \rangle - \langle n_j \rangle \langle n_{j + r} \rangle
\end{equation}
to the Ornstein-Zernike form \cite{Zhang2025}, 
\begin{equation}
        \label{ornstein zernike formula}
    C(r) \sim \frac{e^{-r/\xi}}{\sqrt{r}} ~\cos(kr + \phi_0),  
\end{equation}
where $k, \xi$, and $\phi_0$ are treated as free-fitting parameters. Results are shown in Fig. \ref{floating phase stability}. The resulting fits reveal a clear distinction between the floating and disordered phases. For the floating-phase point, the extracted correlation length is large and comparable to the bulk region size, indicating slowly-decaying, quasi-long-range correlations. In contrast, for a representative point in the disordered phase---obtained from a uniform $L = 121$-site chain at fixed detuning $\delta = 1.59$---the correlations decay rapidly, with a correlation length of only $\sim 5$ sites. This sharp contrast provides a direct diagnostic of the underlying phase: the large correlation length observed in the non-uniform setting confirms that disordered boundary subsystems stabilize genuine floating-phase behavior, rather than producing a disordered state in the bulk region. 

\subsection{2D results}

\subsubsection{Definitions of the square/striated and $(1, 1)$-sublattice OPs}

To identify the square/striated phase, we employ the following OP defined in Ref. \cite{Ebadi2021}: 
\begin{eqnarray}
    \tilde{\mathcal{F}}(k_1, k_2) = (\mathcal{F}(k_1, k_2) + \mathcal{F}(k_2, k_1))/2, \\
    \mathcal{F}(k_1, k_2) = |\displaystyle\sum_{x, y} \exp(i(k_1 x + k_2 y)) \langle \hat{n}_{x, y} \rangle|/N, 
\end{eqnarray}
where $O_{\mathrm{striated}} := \tilde{\mathcal{F}}(\pi, 0) - \tilde{\mathcal{F}}(\pi/2, \pi)$ (here, $N = L_x L_y$ denotes the lattice size). By definition, however, $O_{\mathrm{striated}}$ does not reliably discriminate between the square and striated ordering patterns. We therefore evaluate a second OP defined in Ref. \cite{ORourke2023} and designed to be sensitive exclusively to the striated order: 
\begin{equation}
O_{(1,1)\text{-sublattice}} := \frac{4}{N} \sum_{\substack{x,y \\ x \equiv 1~(\mathrm{mod}\ 2) \\ y \equiv 1~(\mathrm{mod}\ 2)}} \langle \hat{n}_{x,y} \rangle, 
\end{equation}
which quantifies the density deformation on the $(1, 1)$-sublattice.

\subsubsection{Stability of the striated phase at $R_b = 1.6$ with non-uniform boundary detuning}

\begin{figure*}[htbp!]
    \centering
    \includegraphics[width=0.8\textwidth]
    {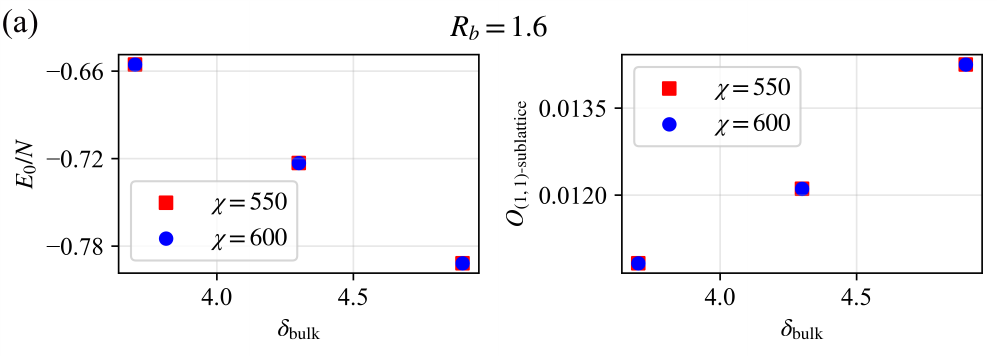}
    \\[1em]
    
    \includegraphics[width=0.5\textwidth]
    {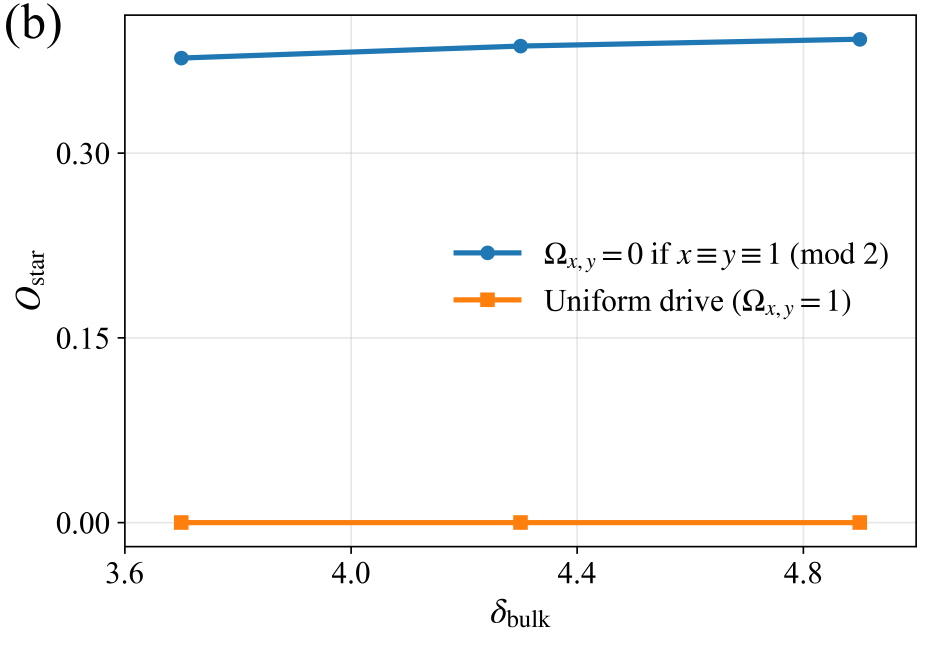}

        
    \caption{%
        \textbf{(a)} Convergence of DMRG ground-state energies and the $(1, 1)$-sublattice order parameter at $R_b = 1.6$ for $\chi = 550$ and $600$. \textbf{(b)} Star order parameter $O_{\mathrm{star}}$ as a function of bulk detuning $\delta_{\mathrm{bulk}}$ for a non-uniform Hamiltonian on a $13 \times 13$ lattice with and without suppression of local density fluctuations on the $(1, 1)$-sublattice. The non-uniform Hamiltonian uses $n_{\mathrm{boundary}} = 4$, $\delta_{\mathrm{boundary}} = 1.8$, and interface detuning mismatch $\alpha = 0.05$.
    }
    \label{striated pattern numerical tests}
\end{figure*}

To further substantiate that the phase stabilized at $R_b = 1.6$ on the $13 \times 13$ array with boundary detuning ramps (main text Fig.~3) corresponds to a striated ordering pattern rather than a pure square-ordered phase, we perform two additional complementary numerical tests: 

\begin{enumerate}
    \item \textit{Bond dimension convergence of weak sublattice excitations.}---The striated phase is characterized by weak but finite excitation density on the $(1, 1)$-sublattice, which could in principle be mistaken for a numerical artifact if insufficient bond dimension is used. To rule out this possibility, we repeat the DMRG simulations at bond dimensions $\chi = 550$ and $600$. In all cases, the ground-state energies agree to nearly the seventh decimal place, and the $(1, 1)$-sublattice order parameter, $O_{(1, 1)-\mathrm{sublattice}}$, remains unchanged within numerical precision (Fig. \ref{striated pattern numerical tests}(a)). This demonstrates that the weak excitations observed on the $(1, 1)$-sublattice are a robust physical feature of the ground state rather than a truncation-induced artifact. 
    \item \textit{Suppression of $(1, 1)$-sublattice density fluctuations.}---As a more stringent diagnostic, we explicitly suppress quantum fluctuations on the $(1, 1)$-sublattice by setting the local Rabi frequency to zero on those sites, 
    \begin{equation}
        \Omega(x, y) = 0; \mod(x, 2) = 1 ~\mathrm{and} \mod(y, 2) = 1 
    \end{equation}
    and $\Omega = 1$ at all other sites. Under this modified Hamiltonian, DMRG simulations with disordered boundary subsystems consistently stabilize the star ground state in the bulk region (Fig. \ref{striated pattern numerical tests}(b)) across the entire range $3.7 \leq \delta_{\mathrm{bulk}} \leq 4.9$. This confirms that the ground state at $R_b=1.6$ for the original non-uniform Hamiltonian (main text Fig.~3) cannot be the square phase, since the stability of the square phase would be unaffected by suppression of density fluctuations on the $(1,1)$-sublattice. This leaves the weakly striated ground state as the only remaining candidate for the ground state at $R_b=1.6$.  
\end{enumerate}

\bibliography{References}